\documentclass{PoS}

\usepackage{cite}
\usepackage{amssymb}
\usepackage{amsmath}

\title{NLO QCD corrections to off-shell $\mathbf{t\bar{t}j}$ production at the LHC}

\ShortTitle{NLO QCD corrections to off-shell $t\bar{t}j$ production at the LHC}

\author{\speaker{Ma\l{}gorzata Worek} \thanks{TTK-16-26}\\
        Institute for Theoretical Particle Physics and Cosmology\\
       RWTH Aachen University, D-52056 Aachen, Germany\\
        E-mail: \email{worek@physik.rwth-aachen.de}}

\abstract{A short summary of results for top-quark pair production in
association with a jet with NLO QCD off-shell effects at the LHC is
given. The calculation is based on the matrix element for the $pp\to
e^+ \nu_e \mu^-\bar{\nu}_\mu b\bar{b} j$ process. All quantum effects
are properly accounted for, which amounts to including all Feynman
diagrams that result in this final state with interference effects as
well as non-zero top-quark width effects. Results for total cross
sections and a few differential distributions at the LHC Run1 energy
of $\sqrt{s}=8$ TeV are presented. In addition, the size of top-quark
off-shell effects is quantified for the total cross section in
presence of rather inclusive cuts on all final states. Numerical
results shown in this proceedings have been obtained with the
\textsc{Helac-NLO} Monte Carlo program. }

\FullConference{Loops and Legs in Quantum Field Theory\\
		24-29 April 2016\\
		Leipzig, Germany}

\begin{document}

%
\section{Introduction}
%

\begin{figure}[t!]
\begin{center}
\includegraphics[width=0.49\textwidth]{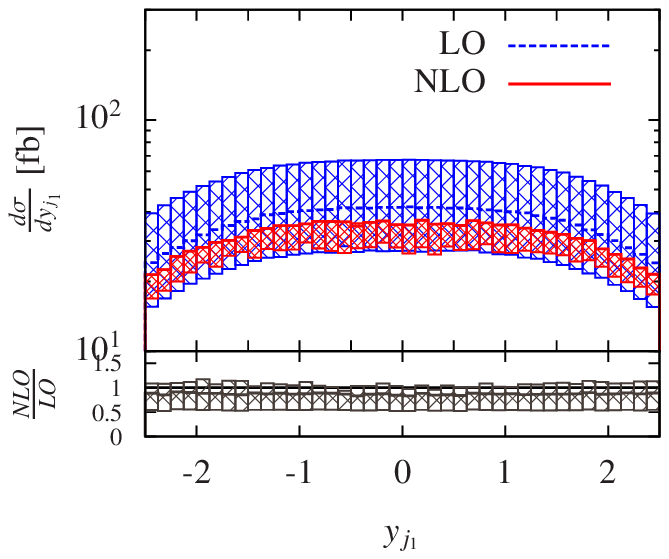}
\includegraphics[width=0.49\textwidth]{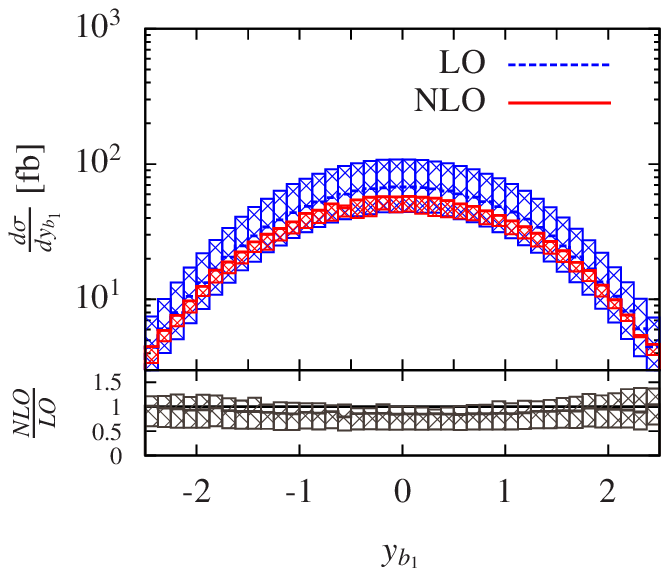}
\end{center}
\caption{\it    Differential cross section distributions as a function
  of  rapidity of the hardest light- and   b-jet for
$pp\to e^+\nu_e \mu^- \bar{\nu}_\mu b\bar{b} j +X$ at the LHC with
$\sqrt{s} = 8 ~{\rm TeV}$.  The uncertainty bands depict the scale
variation. Lower panels display differential K factors and their 
uncertainty bands.}
\label{fig:1}
\end{figure}

Top-quark studies are currently driven by two Large Hadron Collider
(LHC) experiments, i.e. ATLAS and CMS. Besides the determination of
the top-quark mass, $m_t$, and the strong coupling constant,
$\alpha_s$, key measurements include fiducial cross sections, various
infra-red safe differential distributions as well as spin correlations
and top-quark couplings to gauge bosons and the Standard Model (SM)
Higgs boson. In addition, work on constraining parton distribution
functions for the dominant production channel, i.e. the $gg$ channel
is ongoing.  The top-quark phenomenology provides a unique laboratory
where our understanding of the strong interactions, both in the
perturbative and non-perturbative regimes, can be tested. Searches for
rare top-quark decays to probe physics beyond the SM also play a
prominent role at the LHC. Finally, $t\bar{t}$ production is the
dominant background process for studies of the SM Higgs boson and for
new physics searches. The top-quark is an extremely
short-lived resonance and only its decay products can be detected
experimentally. In general, for comparison with data, theoretical
predictions must include top-quark decays. In the SM, the top-quark decays
almost exclusively to a $W$-boson and a $b$-quark. The experimentally
cleanest top-quark decay channel, which comprises leptonic decays of
both $W$-gauge bosons, is called the di-lepton channel. The signature
for this channel consists of two well-isolated and oppositely charged
leptons with high transverse momentum, missing transverse momentum,
$p_T^{miss}$, from invisible neutrinos and two jets that originate
from bottom quarks. However, due to the large collision energy at the
LHC, top-quark pairs are abundantly produced with large energies and
high transverse momenta. Thus, the probability for additional
radiation increases making the $t\bar{t}j$ final state measurable with
high statistics.  To estimate the size of the $t\bar{t}j$ contribution
in the inclusive $t\bar{t}$ sample we show in Table \ref{tab:1} the
cross section for the on-shell $pp\to t\bar{t}j$ production at NLO in
QCD with various $p_T(j)$ cuts on the hard jet. Also shown is its
ratio to the inclusive $pp\to t\bar{t}$ production at NNLO. Both
results are given for the LHC Run2 energy of $13$ TeV and the
top-quark mass of $m_t = 173.2$ GeV. In addition, for parton
distribution functions (PDFs) CT14nlo and CT14nnlo sets have been
employed \cite{Dulat:2015mca}.
%
\begin{table}[h!]
\caption{\label{tab:1} \it The NLO  cross section for the on-shell $pp \to
t\bar{t}j$ production with various $p_T(j)$ cut on the hard jet
and its ratios to $pp\to t\bar{t}$ production at NNLO for the LHC Run2
energy of $13$ TeV.}
\begin{center}
\begin{tabular}{||c||c||c|| }
\hline\hline
 $p_T(j)$ ~~[GeV] &$\sigma^{NLO}_{t\bar{t}j}$ ~~[pb] &
$\sigma^{NLO}_{t\bar{t}j}/\sigma^{NNLO}_{t\bar{t}}$ ~~ $[\%]$\\
\hline
\hline
40 &296.97 $\pm$ 0.29 & 37\\
60 &207.88 $\pm$ 0.19 & 26 \\
80 &152.89 $\pm$ 0.13& 19 \\
100 &115.60 $\pm$ 0.14& 14\\
120 &~~89.05 $\pm$ 0.10& 11\\
\hline\hline
\end{tabular}
 \end{center}
\end{table}
%
NLO results have been obtained with the \textsc{Helac-NLO} package
\cite{Bevilacqua:2011xh}, while the NNLO result,
$\sigma_{t\bar{t}}^{NNLO} = 807$ pb, has been calculated with the
\textsc{Top++} program \cite{Czakon:2011xx}.  For $p_T (j) > 40$ GeV
almost $40\%$ of the top anti-top events are actually accompanied by
an additional hard jet.  Therefore, for a better understanding of
$pp\to t\bar{t}$ kinematics at the LHC it is essential to also study
$pp \to t\bar{t}j$ production in more details. However, $pp\to
t\bar{t}j$ production is important on its own. It can be employed in
the top-quark mass extraction by studying normalised differential
cross sections as a function of the (inverse) invariant mass of the
$t\bar{t}j$ system \cite{Alioli:2013mxa}.  The method has already been
successfully used by experimental groups at the LHC
\cite{Aad:2015waa,CMS:2016khu}. Additionally, $t\bar{t}j$ constitutes
an important background to the SM Higgs boson production in the vector
boson fusion process, $pp\to Hjj \to W^+W^- jj $
\cite{Rainwater:1999sd,Kauer:2000hi}. Indeed, for the $2\ell \,
p_T^{miss} + jets$ final state not $pp \to t\bar{t}$ but the $pp\to
t\bar{t}j$ production process is the dominant background process.  To
understand this better one needs to recall that the VBF signature
consists, among other things, of two tagging jets, denoted as $j^{\,
\, tag}_1$ and $j^{\, \,tag}_2$, that have to be very well separated
in the rapidity plane. In practice, this means that both tagging jets
have to fulfil a large rapidity cut $\Delta y(j^{\, \, tag}_1 j^{\, \,
tag}_2)=|y(j_1^{\, \, tag}) -y(j_2^{\, \, tgg})|> 4$ and they are
required to reside in opposite hemispheres of the detector, i.e.
$y(j_1^{\, \, tag}) \times y(j_2^{\, \, tag}) < 0$. In case of $pp\to
t\bar{t}\to W^+W^-b\bar{b}\to e^+\nu_e \mu^- \bar{\nu}_\mu b\bar{b}$
two tagging jets are $b$-jets. From Figure \ref{fig:1} we can see that
they are centrally distributed in the rapidity plane. Thus, the above
two conditions will dramatically decrease the contribution from this
process. On the other hand, for $pp\to t\bar{t}j\to W^+W^-b\bar{b}j\to
e^+\nu_e \mu^- \bar{\nu}_\mu b\bar{b}j$ production it is sufficient
that only one of the two tagging jets is the $b$-jet since we also
have a light-jet. The latter is distributed uniformly in the rapidity
plane, as can be observed in Figure \ref{fig:1}.  Therefore, in that
case a quite substantial part of the cross section remains causing the
$t\bar{t}j$ production process to dominate among various background
processes. Furthermore, $pp\to t\bar{t}j$ plays a very important role
in searches for physics beyond the SM. For example, it is one of the
main backgrounds to processes such as supersymmetric particle
production \cite{Mangano:2008ha}. Depending on the specific model,
typical signals also include $2\ell^\pm + E_T^{miss}+ jets$, where
$E_T^{miss }$ is due to the escaping lightest supersymmetric
particle. Another exemple comprises the production of top-quark
flavour violating states like for example $M \to \tilde{t}q$, where
$\tilde{t}=t$ or $\bar{t}$. Such states can be singly produced in
association with a top-quark at the LHC.  Thus, the direct signature
will consist of a $tj$ (or $\bar{t}j$) resonance in $t\bar{t}j$ events
\cite{Gresham:2011dg}.
%
%
In view of a correct interpretation of the signals of new physics,
which might be extracted from data, it is of great importance to
understand the background process as precisely as possible.  In this
respect, the need of precision theoretical predictions for $pp\to
t\bar{t}j$ production process is indisputable. NLO QCD corrections for
the $pp\to t\bar{t}j$ production process have been first calculated 
with on-shell top-quarks
\cite{Dittmaier:2007wz,Dittmaier:2008uj}. In the next step decays in
the so-called narrow-width-approximation (NWA) have been included,
first at LO \cite {Melnikov:2010iu} and later on at NLO
\cite{Melnikov:2011qx}. In this approximation various production and
decay stages are treated sequentially. The NWA method allows to
neglect single-resonant and non-resonant amplitude contributions as
well as all interferences and off-shell effects of the top-quark.
This of course leads to significant simplifications in calculations of
higher order corrections for top-quark physics. 
%
\begin{figure}[t!]
\begin{center}
\includegraphics[width=1.0\textwidth]{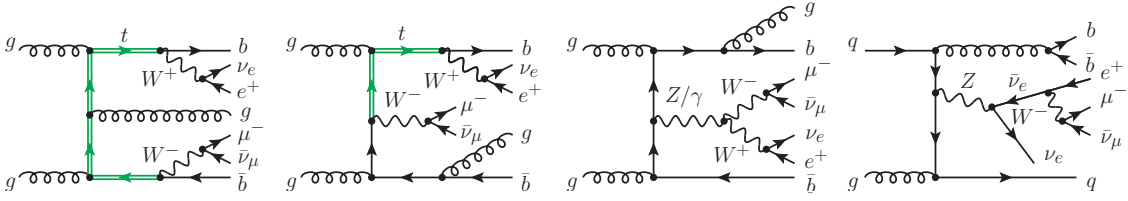}
\end{center}
\caption{\it Representative Feynman diagrams, involving two (first
  diagram), one (second diagram) and no top-quark resonances (third
  diagram),  contributing to the  $pp \to e^+\nu_e \mu^-
  \bar{\nu}_\mu b\bar{b}j$ process at ${\cal O}(\alpha_s^3 \alpha^4)
  $. The last diagram with a single $W$ boson resonance contributes to
  the off-shell  effects of  $W$ gauge bosons.}
\label{fig:2}
\end{figure}
%
Finally, complete NLO QCD corrections to $pp\to t\bar{t}j$ in the
di-lepton channel have been recently provided
\cite{Bevilacqua:2015qha}.  Here the approximation that top quarks are
only produced on-shell has been dropped, and the fully realistic final
state $pp \to e^+ \nu_e \mu^- \bar{\nu}_\mu b\bar{b}j$ has been
considered.  Resonant and non-resonant top quark contributions as well
as all interference effects among them have been consistently taken
into account.  In addition, non-resonant and off-shell effects due to
the finite $W$ gauge boson width were included. A few examples of
Feynman diagrams contributing to the LO process at ${\cal
O}(\alpha_s^3\alpha^4)$ are presented in Figure \ref{fig:2}. Such
calculations are extremely complex and the question remains whether
they are really needed.  Indeed, the size of the top quark off-shell
effects is controlled by the ratio of the top-quark width to the
top-quark mass $\Gamma_t/m_t\approx 10^{-2}$. Thus, the contributions
that are neglected in the NWA should be suppressed at least for
sufficiently inclusive observables like the total cross
sections. Certainly, in case of $pp\to t\bar{t}$ and $pp\to t\bar{t}j$
productions as well as for $pp\to t\bar{t}H$, these effects have been
found to be of the order of $1\% -2\%$ for the total cross section
\cite{Bevilacqua:2015qha,Denner:2010jp,Bevilacqua:2010qb,
Denner:2012yc,Frederix:2013gra,Cascioli:2013wga,Heinrich:2013qaa,
Denner:2015yca}. However, they can be strongly enhanced in exclusive
observables that play an important role in studies of the Higgs boson
and searches for new physics \cite{AlcarazMaestre:2012vp}. Before
presenting a few results for $pp \to e^+ \nu_e \mu^- \bar{\nu}_\mu
b\bar{b}j$ production at ${\cal O}(\alpha_s^4\alpha^4)$ let us shortly
mention here that results for $pp\to t\bar{t}j$ at NLO accuracy
matched with parton showers are also available
\cite{Kardos:2011qa,Alioli:2011as,Czakon:2015cla}.

%
\section{Outline of the Calculation}
%
%

NLO QCD corrections to $pp \to e^+ \nu_e \mu^- \bar{\nu}_\mu
b\bar{b}j+X$ have been calculated with the \textsc{Helac-NLO} Monte
Carlo program, which has already been extensively used for top-quark
related studies at NLO in QCD \cite{Bevilacqua:2009zn,
Bevilacqua:2010ve,Bevilacqua:2011aa,Worek:2011rd,Bevilacqua:2012em,
Bevilacqua:2014qfa}.  \textsc{Helac-NLO} is an extension of the
\textsc{Helac-Phegas} LO MC program \cite{Kanaki:2000ey,
Cafarella:2007pc}, which is based on off-shell Dyson-Schwinger
recursive equations for amplitudes calculation. The
\textsc{Helac-1Loop} program \cite{vanHameren:2009dr}, based on the
Ossola-Papadopoulos-Pittau reduction technique \cite{Ossola:2006us}
and the reduction code \textsc{CutTools} \cite{Ossola:2007ax}, has
been used to deal with virtual corrections.  For the evaluation of the
scalar integrals the \textsc{OneLOop} library has been employed
\cite{vanHameren:2010cp}.  Singularities from soft or collinear parton
emission have been isolated via subtraction methods for NLO
QCD. Specifically, two methods have been made use of, the commonly
used Catani-Seymour dipole subtraction \cite{Catani:2002hc}, and the
Nagy-Soper subtraction scheme \cite{Bevilacqua:2013iha}, both
implemented in the \textsc{Helac-Dipoles} software
\cite{Czakon:2009ss}. Phase space integration has been executed with
the help of \textsc{Kaleu} \cite{vanHameren:2010gg}. The process under
consideration requires a special treatment of unstable top-quarks.  We
regularize intermediate top-quark resonances in a gauge-invariant way
via the complex-mass-scheme \cite{Denner:2005fg}. In this approach the
top-quark width $\Gamma_t$ is incorporated into the definition of the
(squared) top-quark mass through $\mu^2_t=m^2_t -im_t \Gamma_t$.  As a
consequence top-quark contributions are consistently described by
Breit-Wigner distributions at the cost of evaluating all matrix
elements with complex top-quark masses.

%
\section{Numerical Results for the LHC}
%
%

\begin{figure}[t!]
\begin{center}
\includegraphics[width=0.49\textwidth]{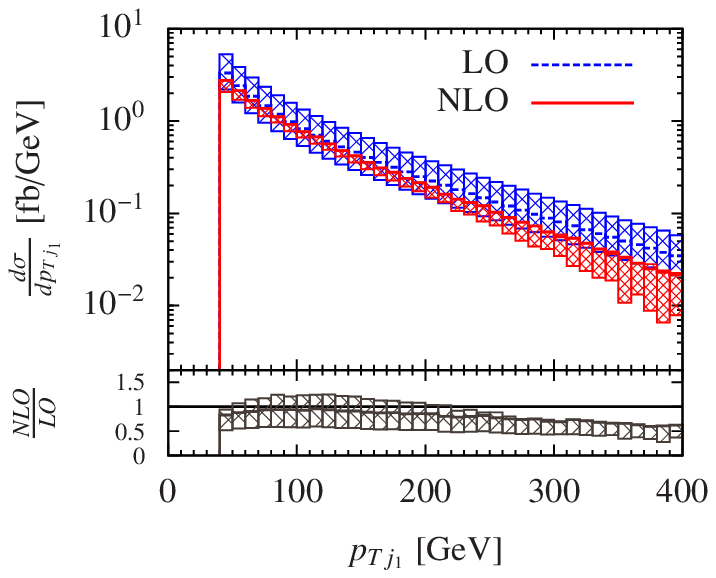}
\includegraphics[width=0.49\textwidth]{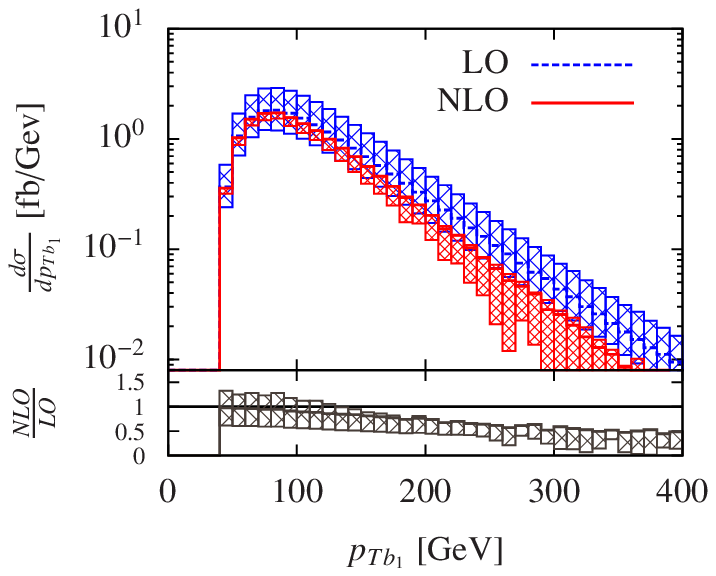}
\includegraphics[width=0.49\textwidth]{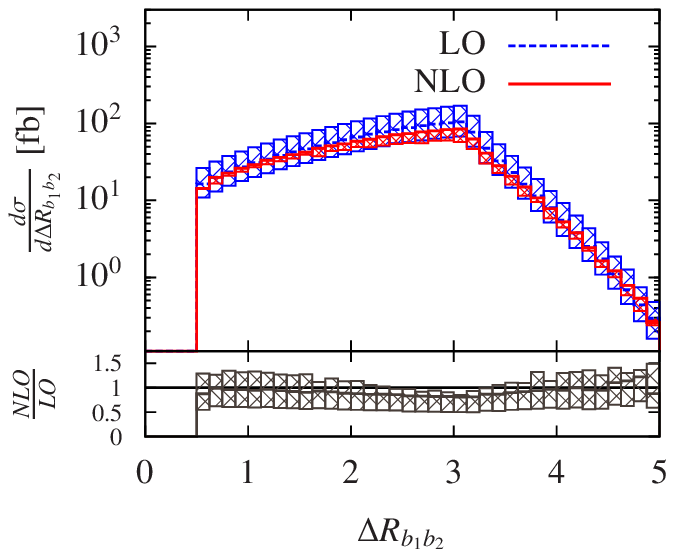}
\includegraphics[width=0.49\textwidth]{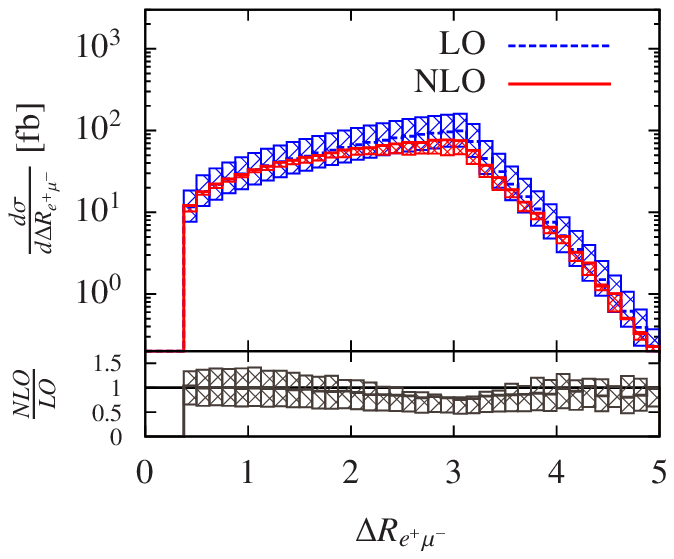}
\includegraphics[width=0.49\textwidth]{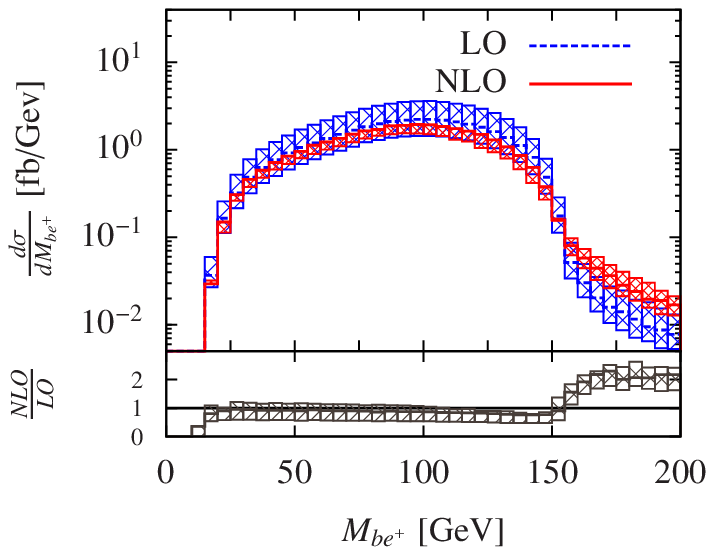}
\includegraphics[width=0.49\textwidth]{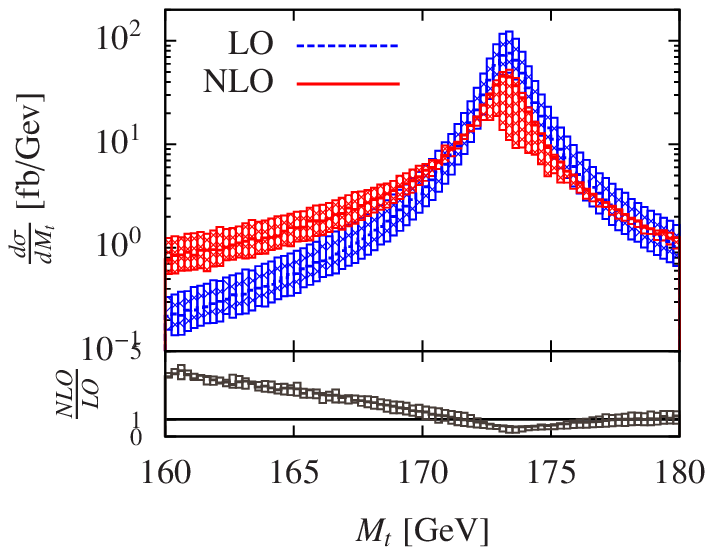}
\end{center}
\caption{\it   Differential cross section distributions as a function
of $p_{T}(j_1)$, $p_{T}(b_1)$, $\Delta R_{b_1b_2}$, $\Delta
R_{e^+\mu^-}$, $M_{be^+}$, and $M_{t}$ for $pp\to e^+\nu_e \mu^-
\bar{\nu}_\mu b\bar{b} j +X$ at the LHC with $\sqrt{s} = 8 ~{\rm
TeV}$.  The uncertainty bands depict the scale variation.  Lower
panels display differential K factors and their uncertainty bands.}
\label{fig:4}
\end{figure}

In the following we present predictions for the total cross section and a
few differential observables at the LHC for the collider energy of $8$
TeV. The input parameters used in our calculations are given in Table
\ref{tab:2}.
%
\begin{table}[h!]
\caption{\label{tab:2} \it SM parameters in the $G_\mu$ scheme.}
\begin{center}
\begin{tabular}{||c||c||}
\hline\hline
 $G_{F}=1.16637 \cdot 10^{-5} ~{\rm GeV}^{-2}$ & 
$m_{t} =173.3    ~{\rm GeV}$   \cite{ATLAS:2014wva}\\
$m_{W}=80.399 ~{\rm GeV}$ & $\Gamma_{W} 
= 2.09974 ~{\rm  GeV}$ \\
$m_{Z}=91.1876  ~{\rm GeV}$
&$\Gamma_{\rm Z} = 2.50966 ~{\rm  GeV}$ \\
$\Gamma_{t}^{LO} = 1.48132 ~{\rm GeV}$  \cite{Jezabek:1988iv}&
$\Gamma_{t}^{NLO} = 1.3542 ~{\rm  GeV}$ \cite{Jezabek:1988iv}
\\
\hline\hline
\end{tabular}
 \end{center}
\end{table}
%
We employ the MSTW2008NLO (LO) PDFs \cite{Martin:2009iq} and use the
running of the strong coupling constant $\alpha_s$ with two-loop
(one-loop) accuracy as provided by the LHAPDF library. Suppressed
contributions induced by the bottom-quark parton density are
neglected. All final-state partons with pseudorapidity $|\eta| < 5$
are recombined into jets via the infrared safe anti-$k_T$ jet
algorithm \cite{Cacciari:2008gp} where a cone size of $R = 0.5$ has
been chosen. Additional cuts are imposed on the transverse momenta and
the rapidity of recombined jets as well as the distance between jets:
\begin{equation}
\label{eq1}
p_{T}(j)>40 ~{\rm GeV}\,, \quad\quad  |y(j)|<2.5\,, \quad \quad
 \Delta R_{jj}>0.5 \,.
\end{equation}
Basic selection is also applied to decay products of top-quarks: 
\begin{equation}
\label{eq2}
p_{T}(\ell)>30 ~{\rm GeV}\,,  \quad \quad p^{miss}_{T} >40 ~{\rm GeV}\,,
\quad \quad 
\Delta R_{\ell\ell}>0.4\,,  \quad \quad \Delta R_{\ell j}>0.4\,, \quad
\quad
|y(\ell)|<2.5\,.
\end{equation}
The factorisation and renormalisation  scales,
$\mu_F$  and $\mu_R$  respectively, have
been set to a common value $\mu_0=m_t$. We start with a
discussion of the total cross section. At the central scale, we have
obtained:
\begin{equation}
\begin{split}
\sigma^{ LO}_{pp\to e^+\nu_e\mu^-\bar{\nu}_\mu b\bar{b}j}
&= 183.1^{+112.2 \, (61\%)}_{-64.2 \, (35\%)} ~{\rm fb}\,,
\\
\sigma^{NLO}_{pp\to e^+\nu_e\mu^-\bar{\nu}_\mu b\bar{b}j}
&= 159.7^{-33.1 \,(21\%)}_{-7.9 \, (5\%)} ~ {\rm
  fb}\,.
\end{split}
\end{equation}
Thus, the ${\cal K}$ factor is $0.87$. Additionally, the inclusion of the NLO
corrections reduces the scale dependence from $61\%$ to $21\%$. 
In the next step we present
various differential distributions obtained by applying the cuts
specified in Eq.~(\ref{eq1}) and Eq.~(\ref{eq2}). In Figure
\ref{fig:4} differential cross section distributions as function of
the transverse momentum of the hardest (in $p_T$) light- and $b$-jet
are shown. The upper panels display absolute LO (the blue-dashed
curve) and NLO (the red-solid curve) predictions together with
corresponding uncertainty bands, which are calculated as maximum and
minimum out of the following set $\left\{0.5 m_t, m_t,
2m_t\right\}$. The lower panels display the differential ${\cal K}$
factor. Higher order corrections do not simply rescale the shape of
the LO distributions. Instead distortions of the order
of $ - 50\%$ are visible within the plotted range. Negative NLO
corrections in $p_T$ tails simply mean that the LO result is higher
than the NLO one. Moreover, we note that the NLO error bands do not
fit within the LO ones as one should expect from
a well-behaved perturbative expansion. This situation can be cured
with a judicious choice of a dynamic scale. The dynamic scale
should take into account $p_T$ of  top-quark decay products so its
value would increase in the tails of $p_T$ distributions. On the other
hand, close to the $t\bar{t}$ treshold, it should reduce to the
top-quark mass since there $\mu_0=m_t$ behaves  correctly. On the
contrary, for dimensionless distributions  moderate and quite
stable NLO corrections are obtained. As an example in Figure \ref{fig:1}
differential cross section  distributions as a function of the
rapidity of the hardest light- and $b$-jet are given, whereas in
Figure \ref{fig:4} differential cross section  distributions as a
function of $\Delta R_{b_1b_2}= \sqrt{\Delta \phi^2_{b_1b_2} + \Delta
  y^2_{b_1b_2}}$  and $R_{e^+\mu^-}=\sqrt{\Delta \phi^2_{e^+\mu^-} +
  \Delta y^2_{e^+\mu^-}}$ are provided. 
%
%
These angular distributions receive contributions from all scales,
most notably from those that are sensitive to the threshold for the
$t\bar{t}j$ production. For our scale choice, effects of the phase
space regions close to this threshold dominate and a dynamic scale
will not alter this behaviour. And finally, in Figure \ref{fig:4} the
differential cross section distribution as a function of the invariant
mass of the positron and the $b$-jet, $M_{be^+}$, is shown. Also given
there is the differential cross section distribution as a function of
the invariant mass of the top-quark in the vicinity of its resonance,
$M_t$. Both observables are particularly sensitive to off-shell
top-quark effects.  We will concentrate first on the $M_{be^+}$
distribution, where the $e^+b$ pair is formed by selecting the $b$-jet that
yields the smallest invariant mass. In the LO NWA this
observable is characterized by a sharp upper bound
\begin{equation}
M_{be^+} = \sqrt{m_t^2 -m_W^2} \approx 153 ~{\rm GeV}\,.
\end{equation}
Additional radiation and off-shell top-quark effects as well as
off-shell effects of $W$ gauge bosons introduce a smearing to the
region, which is highly sensitive to the details of the description of
the process. In the case of $t\bar{t}$ production, this observable has
proved to be particularly important for the extraction of the
top-quark mass with a very high precision
\cite{Chatrchyan:2013boa,Aad:2015nba}.  The top quark mass can be
determined either from the shape of the distribution away from the
kinematical endpoint or from the behavior of the observable in the
vicinity of that point. In the region below $153$ GeV, the NLO
corrections to $M_{be^+}$ are moderate, however, above the kinematic
upper bound they can be as high as $100\%$. Even though the
contribution to the total cross section from this region is fairly
small, the impact on the top-mass measurement might be non-negligible.
Higher order corrections and off-shell contributions impact
greatly another observable that is highly sensitive to the top-quark
mass extraction, i.e. the invariant mass of the top quark,
$M_t=M_{b\ell \nu_\ell }$. In the presence of off-shell top-quark
contributions, a clear Breit-Wigner structure is visible. We can 
see that below the top-quark resonance the size of NLO
corrections is huge due to final-state gluon radiation that is not
recombined with the top-quark decay products. Once the kinematical
region above  $M_t \sim m_t$ is reached NLO corrections become
moderate. Since top quark mass measurements are carried out using the
$M_t$ distribution precise theoretical modeling of top-quark decays is
of paramount importance. When aiming at the $m_t$
extraction with a precision of $1$ GeV or less, theoretical predictions that
are used in such analyses should go beyond simple approximation of
factorising top-quark  production and decays.

In summary, NLO QCD corrections to the $pp\to
e^+\nu_e\mu^-\bar{\nu}_\mu b\bar{b}j +X$ production process have been
calculated. We have shown that they are moderate for the total cross
section, i.e.  of the order of $13\%$.  Nevertheless, their impact on
some differential distributions is much larger. A special case are
differential observables that are used in the top-quark mass
extraction.  Shape-based $m_t$ measurements depend on precise modeling
of differential distributions, thus, top-quark decays in NWA are
simply not good enough.

%
\section*{Acknowledgement}
%
%

The author would like to thank the organizers of the $13^{th}$
Workshop on Loops and Legs in Quantum Field Theory for a kind
invitation and very pleasant atmosphere during the conference. The
work was supported by the DFG under Grant: {\it "Signals and
Backgrounds Beyond Leading Order. Phenomenological studies for the
LHC".}

\end{document}